# Changes in chromatin state in donors subjected to physical stress

Yuriy Shckorbatov[1]*, Valeriy Samokhvalov[2], Dariya Bevziuk[2], Maxim Kovaliov[2]

[1]*Institute of Biology, Kharkiv National University, 61077, Kharkiv, Ukraine*
*E-mail address: yury.g.shckorbatov@univer.kharkov.ua*

[2]*Kharkiv National Medical University, 61022, Kharkiv, Ukraine*

The purpose of the present study is to evaluate changes in chromatin of human buccal epithelium under the influence of stressing factor - dosed physical activity. Investigations were performed in a group of students (13 men) of age 19-23. Cells were stained on a slide by a 2% orcein solution in 45% acetic acid during 1 h. The following physiological indexes were determined: arterial blood pressure, pulse frequency, and frequency of breathing. The physical stress produced by the dosed physical activity causes the considerable increase of degree of heterochromatinization in the cell nuclei of human buccal epithelium. As a rule, the level of heterochromatinization increases after first stage of training, but in some donors it increases significantly only after the second stage of training.

## 1. Introduction

Studies of biological mechanisms of stress are of considerable importance. The physiological concept of stress at organism level was created in 1930-th by H. Selye [1]. Changes in the state of cells under the action of different factors of chemical and physical nature, regarded as factors of stress, were described in works of D. Nasonov [2].

Reactions of organism to external influences at cellular and organism levels are closely associated. A. Policard was one of the first who paid attention to cellular basis of organism pathologies [3]. It is well known that membrane permeability of erythrocytes changes in people with thalassemia [4]. The studies of cellular mechanisms of pathology are necessary to reveal mechanisms of different diseases [5,6]. In this regard, cytological research in the area of physiology of stress is of great interest. The connections between donor's age and state of chromatin [7] and between fatigue and state of chromatin [8] were shown in buccal epithelium cells. The purpose of the present study is to evaluate changes in chromatin of human buccal epithelium under the influence of stressing factor - dosed physical activity.

## 2. Materials and methods

Investigations were performed in a group of students of Kharkov Medical University (13 men) of age 19-23. Students were divided in two groups, investigated in different days: group 1 (8 students) and group 2 (5 students). Sport exercises on stationary training bicycle VE02 (Kiev) of power 400 W, 1 min were realized twice with a period of rest 30 min. The following physiological indexes were determined: arterial blood pressure by Korotkov's method with the use of manometer of OTK H87, pulse frequency by pinning of radial artery on a wrist, and frequency of breathing by the standard count. All indexes, and also the test on buccal epithelium cells were measured before physical activity and directly after physical activity.

The cells of buccal epithelium were obtained from the internal surface of cheek by scraping with blunt spatula and processed as described in [9]. Cells were placed in buffer solution of the following composition: 3,03 mM phosphate buffer (pH=7,0) with addition of a 2,89 mM calcium chloride. Cells were stained on a slide by a 2% orcein solution in 45% acetic acid during 1 h. State of chromatin in cells was analyzed at magnification x 600. The number of granules of heterochromatin was counted in one nucleus and the average heterochromatin granule quantity (HGQ) was determined for 30 nuclei. Statistical processing of experimental data was realized by Student's method. Standard error did not exceed a 5% of the measured HGQ value.

## 3. Results

Physical activity induced HGQ increase in both investigated groups (Fig. 1, 2 and Tables 1, 2 of Suppl. Materials). In group 1 after the first minute of physical activity was registered a significant increase of HGQ was registered in 5 donors of 8 and after second training period HGQ increased in cells of 1 donor. No increase of HGQ was observed in two donors. In group 2 the growth of HGQ was observed in all donors, but in cells of donor P the significant increase of HGQ was observed only at the second stage of training. The mean data for two groups (13 donors) are presented in Fig. 3 and Table 3 of Suppl. Materials. As one can see, these data prove the conclusions made for individual cases.

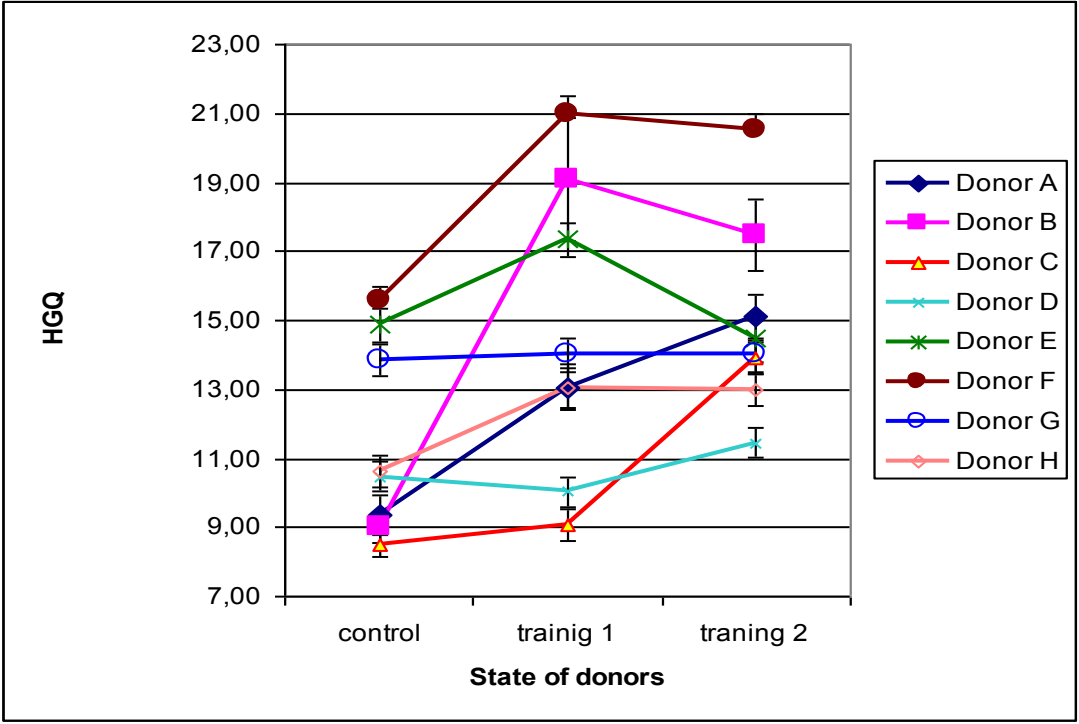

**Fig. 1.** Influence of the dosed physical activity on the state of chromatin in buccal epithelium cells (experiment 1).

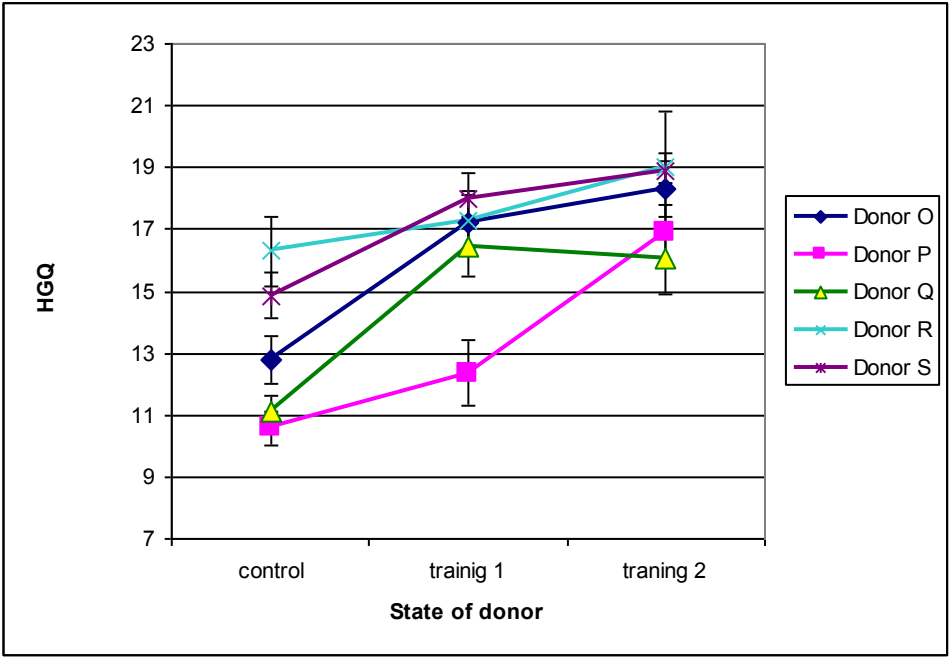

**Fig. 2.** Influence of the dosed physical activity on the state of chromatin in buccal epithelium cells (experiment 2).

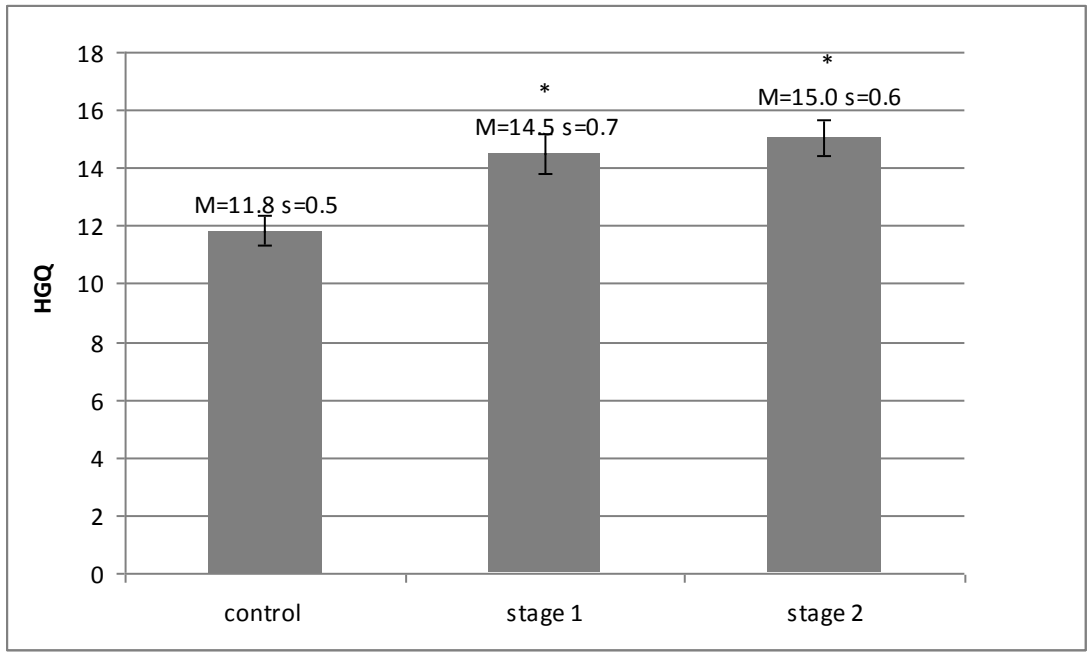

**Fig. 3. Influence of the dosed physical activity on the state of chromatin in buccal epithelium cells. Mean data for two groups of donors (13 donors).**

In parallel with HGQ investigation we analyzed the influence of the physical activity on some physiological parameters (Table 4 of Suppl. Materials). As one can see from the presented data, the physical activity caused the considerable increase of indexes of arterial blood pressure, pulse frequency and breathing frequency that is related to activating of the sympato-adrenal system and forming of the first phase of stress by Selye [1].

## 4. Discussion

In this work we observed the increase of HGQ and consequently the heterochromatinization under the influence of physical activity. Our donors experienced a certain physical stress that manifested itself in an increased systolic pressure, diastolic pressure, pulse pressure, breathing frequency (see Table 4). We suppose that pronounced effect of the first period of physical activity is connected with stimulating effect of physical activity on sympato-adrenal system. We observed the increase of HGQ in 11 donors out of 13. However the significant increase was observed on the different stages of training. In 11 donors it happened after the first stage of training, and in 3 – after the second stage. As a rule, if at the first stage of training HGQ increased significantly after the second stage of training the level of heterochromatinization not increased significantly or even decreased, so no cumulative effect of training was observed. We did not observe HGQ increase in 2 donors (donors D and G). We speculate that differences in reaction to training are connected with differences in physical state of donors (the D and G donors were very well trained, which could result in the lack of response to a moderate physical activity).

Previously it was shown that in donors subjected to the physical activity (walking with weights) the number of heterochromatin granules in cells of buccal epithelium increased [8]. In the cited work the assessment of the level of activity could be made only approximately. In the present work assessment of energy loss was made more accurately that makes it possible to evaluate the influence of physical stress more objectively.

An increase of HGQ is typical for cell responses to various harmful factors (heat, poisons, UV irradiation, microwave radiation [10] i.e. cell stress. We suspect, that granules we observed may be related to stress granules. The phenomenon of formation of so-called stress granules in interphase nuclei under the influence of heat shock was first shown in 1997 [11, 12].

We believe that the process of heterochromatinization in this research is connected with increase of the level of stress hormones. The previously observed *in vitro* heterochromatinization under the influence of adrenaline, noradrenalin and hydrocortisone supports this view [13].

## 5. Conclusions

1. The physical stress produced by the dosed physical activity causes the considerable increase of degree of heterochromatinization in the nuclei of cells of human buccal epithelium.
2. As a rule, the level of heterochromatinization increases after first stage of training, but in some donors it increases significantly only after the second stage of training.

**Supplementary materials**

**Table 1. Influence of the dosed physical activity on the state of chromatin in buccal epithelium cells (experiment 1).**

| Donors | Control | | Training 1 | | Training 2 | |
|---|---|---|---|---|---|---|
| | HGQ | Standard error | HGQ | Standard error | HGQ | Standard error |
| Donor A | 9,35 | 0,59 | 13,07 | 0,64 | 15,11 | 0,67 |
| Donor B | 9,00 | 0,44 | 19,09 | 1,76 | 17,47 | 1,04 |
| Donor C | 8,50 | 0,35 | 9,07 | 0,48 | 13,90 | 0,46 |
| Donor D | 10,47 | 0,44 | 10,03 | 0,44 | 11,46 | 0,42 |
| Donor E | 14,87 | 0,49 | 17,33 | 0,47 | 14,50 | 0,68 |
| Donor F | 15,57 | 0,44 | 21,00 | 0,53 | 20,55 | 0,45 |
| Donor G | 13,83 | 0,46 | 14,00 | 0,48 | 14,00 | 0,48 |
| Donor H | 10,63 | 0,48 | 13,07 | 0,58 | 13,00 | 0,45 |

**Table 2. Influence of the dosed physical activity on the state of chromatin in buccal epithelium cells (experiment 2).**

| Donors | Control | | Training 1 | | Training 2 | |
|---|---|---|---|---|---|---|
| | HGQ | Standard error | HGQ | Standard error | HGQ | Standard error |
| Donor O | 12,80 | 0,59 | 17,21 | 0,89 | 18,33 | 0,91 |
| Donor P | 10,57 | 0,44 | 12,36 | 1,06 | 16,89 | 0,93 |
| Donor Q | 11,14 | 0,35 | 16,44 | 0,99 | 16,03 | 1,15 |
| Donor R | 16,29 | 0,44 | 17,31 | 0,94 | 19,00 | 0,48 |
| Donor S | 14,87 | 0,49 | 18,00 | 0,85 | 18,91 | 1,88 |
| Donor O | 12,80 | 0,59 | 17,21 | 0,89 | 18,33 | 0,91 |

**Table 3. Influence of the dosed physical activity on the state of chromatin in buccal epithelium cells. Mean data for two groups of donors (13 donors).**

|            | HGQ   | Standard error |
|------------|-------|----------------|
| Control    | 12,00 | 0,65           |
| Training 1 | 14,86 | 0,89           |
| Training 2 | 16,03 | 0,67           |

**Table 4. Influence of the dosed physical activity on the state of physiological indexes. The mean data for two groups of donors (13 donors)**

| Index | Mean value ± S.E. |
|---|---|
| Systolic pressure in control | 120,7±3,5 |
| Systolic pressure after the first period of activity | 145,0±4,5* |
| Systolic pressure after the second period of activity | 146,2±4,5* |
| Diastolic pressure in control | 72,6±1,9 |
| Diastolic pressure after the first period of activity | 60,4±2,2* |
| Diastolic pressure after the second period of activity | 58,8±2,3* |
| Pulse pressure in control | 48,1±3,4 |
| Pulse pressure after the first period of activity | 84,6±5,1* |
| Pulse pressure after the second period of activity | 87,3±5,4* |
| Pulse in control | 81,4±5,8 |
| Pulse after the first period of activity | 145,2±7,3* |
| Pulse after the second period of activity | 147,6±6,4* |
| Breathing frequency in control | 19,6±1,4 |
| Breathing frequency after the first period of activity | 27,8±2,2* |
| Breathing frequency after the second period of activity | 33,2±2,1* |

* $P<0.05$ by comparison to control